\documentclass[aps,prd,secnumarabic,nobibnotes,twocolumn,superscriptaddress]{revtex4-1}
\usepackage{amsfonts}
\usepackage{mathrsfs}
\usepackage{amsmath}
\usepackage{color}
\usepackage{natbib}
\usepackage{graphicx}
\usepackage{bm}
\usepackage{amssymb}
\usepackage{xspace}
\usepackage{epstopdf}
\usepackage{dcolumn}
\usepackage{multirow}
\usepackage[colorlinks=true, letterpaper=true, pdfstartview=FitV, linkcolor=blue, citecolor=blue, urlcolor=blue]{hyperref}
\usepackage{wrapfig}

\makeatletter

\newcommand{\Rmnum}[1]{\expandafter\@slowromancap\romannumeral #1@}
\makeatother

\begin{document}
\title{Structure, phase stability, half-metallicity, and fully spin-polarized Weyl states in compound NaV$_2$O$_4$: a new example for topological spintronic material}

\author{Tingli He}
\affiliation{School of Materials Science and Engineering, Hebei University of Technology, Tianjin 300130, China.}

\author{Xiaoming Zhang}
\email{zhangxiaoming87@hebut.edu.cn}
\affiliation{School of Materials Science and Engineering, Hebei University of Technology, Tianjin 300130, China.}

\author{Tie Yang}
\affiliation{School of Physical Science and Technology, Southwest University, Chongqing 400715, China.}

\author{Ying Liu}
\affiliation{School of Materials Science and Engineering, Hebei University of Technology, Tianjin 300130, China.}

\author{Xuefang Dai}
\affiliation{School of Materials Science and Engineering, Hebei University of Technology, Tianjin 300130, China.}

\author{Guodong Liu}
\email{gdliu1978@126.com}
\affiliation{School of Materials Science and Engineering, Hebei University of Technology, Tianjin 300130, China.}

\begin{abstract}
Here, we systematically investigate the structure, phase stability, half-metallicity, and topological electronic structure for a new topological spintronic material NaV$_2$O$_4$. The material has a tetragonal structure with excellent dynamical and thermal stabilities. It shows a half-metallic ground state, where only the spin-up bands present near the Fermi level. These bands form a Weyl nodal line close to the Fermi level, locating in the $k_z$ = 0 plane. The nodal line is robust against SOC, under the protection of the mirror symmetry. The nodal line band structure is very clean, thus the drumhead surface states can be clearly identified. Remarkably, the nodal line and drumhead surface states have the 100$\%$ spin polarization, which are highly desirable for spintronics applications. In addition, by shifting the magnetic field in-plane, we find that the Weyl nodal line can transform into single pair of Weyl nodes. The Weyl-line and Weyl-node fermions in the bulk, as well as the drumhead fermions on the surface are all fully spin-polarized, which may generate new physical properties and promising applications.
\end{abstract}
\maketitle

\section{INTRODUCTION}
The discoveries of Dirac and Weyl semimetal with the nontrivial band topology have sparked a revolution in condensed matter physics~\cite{1,2,3,4,5,6,7,8,9,10}. The unique linear band crossing in these topological semimetals will cause a wide range of exotic transport and optical phenomena~\cite{11,12,13,14}. According to different band crossings in the electronic structure, topological semimetals can be classified into several groups such as Weyl/Dirac nodal line semimetals~\cite{17,18,19}, semimetals~\cite{20,21,22,23,24,25,26,27}, nodal-chain semimetals~\cite{28,29}, and nodal surface semimetals~\cite{30,31,32}. Among them, nodal line semimetal has attracted particular interests because it can be seen as the parent state of various quantum states, such as the topological insulator and Weyl semimetal states~\cite{33,34}. Previously, the studies on nodal line semimetals were mainly conducted in the absence of spin-orbit coupling (SOC)~\cite{35,36}, under the protection of the mirror symmetry or the combing of the time reversal and the inversion symmetries.

In magnetic topological materials, the magnetic ordering breaks the time reversal symmetry, so additional symmetry is required to protect the topological state. Historically, pyrochlore iridate~\cite{37} and half metal HgCr$_2$Se$_4$~\cite{38}, which host several pairs of time-reversal-breaking Weyl nodes, are the earliest proposal of magnetic topological semimetals. Later, more magnetic Weyl semimetals have been proposed, including the stacking Kagome Lattice Mn$_3$Sn/Ge~\cite{39} and Heusler compounds such as Co$_2$TiX (X = Si, Ge, or Sn) ~\cite{40,41}, etc. Soon after, great progress has been made in magnetic Dirac semimetals, and several candidates in CuMnAs~\cite{42}, orthorhombic LaMnO$_3$~\cite{43}, MnF$_3$~\cite{44} and EdCd$_2$As$_2$~\cite{45} have been proposed. Very recently, it has seen few reports on the magnetic nodal line semimetals (NLSMs) in three-dimensional materials Fe$_3$GeTe$_2$~\cite{46}, $\beta$-V$_2$PO$_5$~\cite{47} and Li$_3$(FeO$_3$)$_2$~\cite{48}, and two-dimensional ones LaCl~\cite{49,50}, MnN~\cite{51}, and CrAs$_2$~\cite{52}. The spin-polarization in magnetic topological semimetals makes the topological electrons potential for spin manipulation and spintronics applications. For this consideration, topological half-metals are the most desirable, because their fermionic states can have a 100$\%$ spin-polarization. However, candidate materials for topological half-metals, especially for nodal line half-metals, are quite limited currently.

In this work, we report an ideal nodal-line half-metal state in NaV$_2$O$_4$ compound. Its topological band structure shows following features: (i) the band crossing forms one Weyl nodal line near the Fermi level; (ii) the nodal line has a 100$\%$ spin-polarization because it arise from the states in single spin channel; (iii) the nodal line can survive under SOC, protected by symmetry; (iv) the nodal line is robust against lattice strain and the electron correlation effects; (v) the nodal line has a large linear energy region and shows definite spin-polarized drumhead surface states. In addition, we show that the nodal line can transform into single pair of Weyl nodes by shifting the magnetization into in-plane. The work provides an excellent material to investigate the novel properties of fully spin-polarized fermionic states with time-reversal breaking.

\section{METHODS}

In this work, the first-principles calculations were performed by using the Vienna ab initio Simulation Package ~\cite{53}. The exchange-correlation potential was adopted by the generalized gradient approximation (GGA) of Perdew-Burke-Ernzerhof (PBE) functional~\cite{54}. The cutoff energy was set as 500 eV. The Brillouin zone was sampled by a Monkhorst-Pack~\cite{55} $k$-mesh with size of 15 $\times$ 15 $\times$ 15. During our calculations, we applied the GGA + $U$ method to account for the Coulomb interaction of 3d orbitals of V atom ~\cite{56,57}. The effective $U$ value for V was chosen as 4 eV. To be noted, shift the $U$ values will not change the conclusion of our work. For lattice optimization, the self-consistent field convergence for the total energy and the force variation were set as $10^{-7}$  eV and 0.001 eV/\AA, respectively. The phonon spectrum are calculated by using the Nanodcal Package~\cite{58}. The topological surface states are calculated based on the maximum-localized Wannier functions, by using the WANNIERTOOLS package~\cite{59,60}.

\section{ CRYSTAL AND MAGNETIC STRUCTURES}

\begin{figure}
\includegraphics[width=8.8cm]{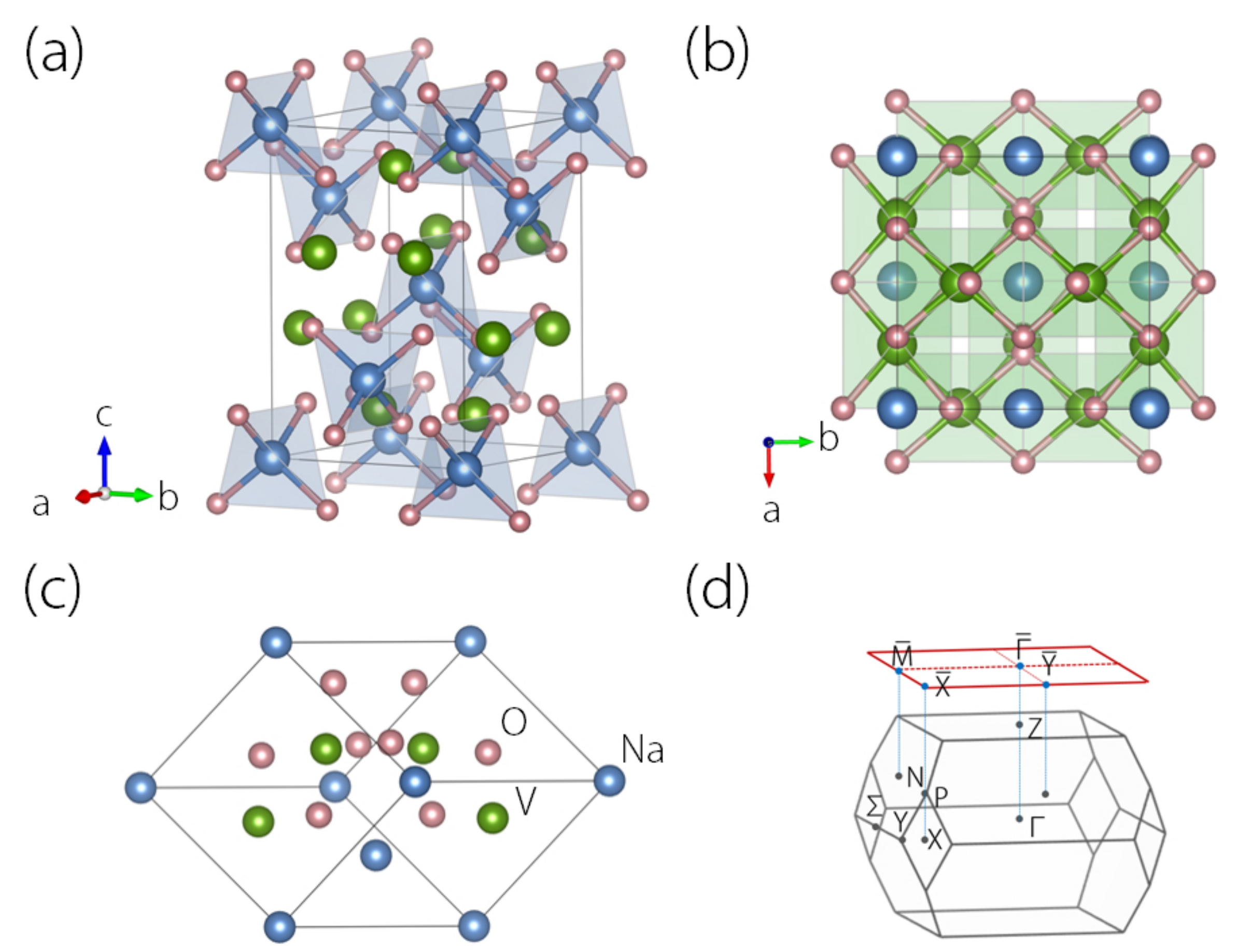}
\caption{(a) Side and (b) top views of the conventional cell of NaV$_2$O$_4$ compound. (c) The primitive cell and (d) the bulk Brillouin zone with the corresponding (001) surface Brillouin zone of NaV$_2$O$_4$ compound.
\label{fig1}}
\end{figure}

The compound NaV$_2$O$_4$ has a tetragonal structure with the space group $I4_1$/$amd$ (No. 141) and the point group of $D_{4h}$. The crystal structure is shown in Fig.~\ref{fig1}(a) and (b). We can find that, the fundamental blocks of the Na-V-O system consist of the stacks between the NaO$_4$ tetrahedra and VO$_6$ octahedra. In the structure, the Na and V atoms locate at the $4b$ (0.000, 0.250, 0.375) and the $8c$ (0.000, 0.000, 0.000) Wyckoff sites, respectively. The O atoms occupy the $16h$ Wyckoff sites (0.000, $\mu$, $\nu$ ). Figure~\ref{fig1}(c) shows the primitive cell form of NaV$_2$O$_4$ compound. One conventional unit cell contains two units of such primitive cell. The structure of NaV$_2$O$_4$ was initially proposed by the Materials Project~\cite{61}, which was demonstrated to be energetically and dynamically stable. In this work, the structure of NaV$_2$O$_4$ compound has been fully relaxed. The optimized lattice values are a = b = 6.053 \AA, c = 9.008 \AA, $\mu$ = 0.036, and $\nu$ = 0.772.

\begin{figure}
\includegraphics[width=8.8cm]{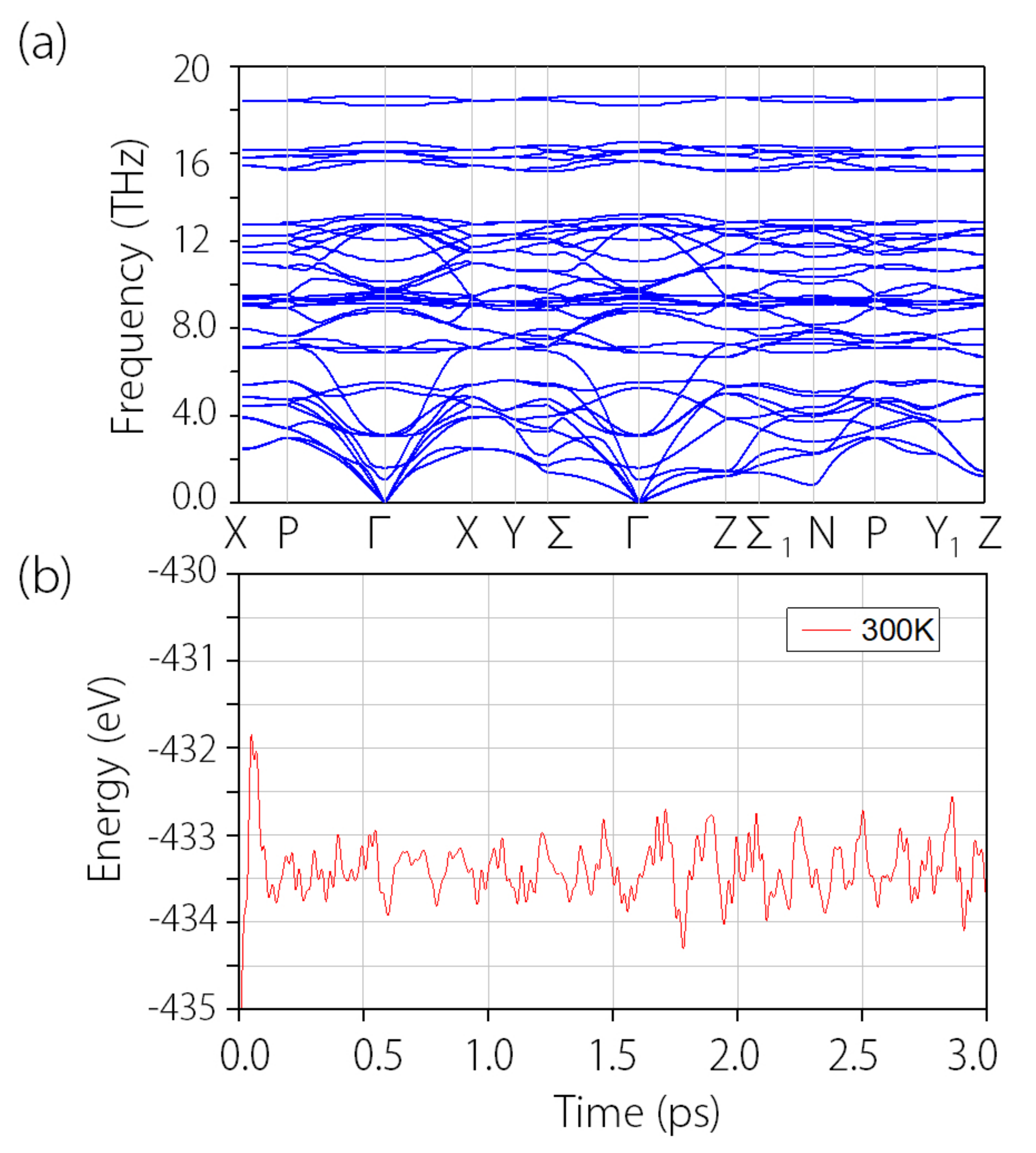}
\caption{The calculated phonon spectrum (a) and the total potential energy fluctuation (b) of NaV$_2$O$_4$ during the AIMD simulation at 300 K.
\label{fig2}}
\end{figure}

Here, we check the stability of the obtained structure. In Fig.~\ref{fig2}(a), we show the phonon spectrum of NaV$_2$O$_4$ compound. We can find that the material has a nice dynamical stablity with no imaginary modes in the whole Brillouin zone. We have also performed the ab initio molecular dynamics (AIMD) simulations to estimate the thermal stability of the obtained structure. The evolution of free energy for NaV$_2$O$_4$ compound at 300K during the AIMD simulation is shown in Fig. 2(b). After running 2000 steps (3 ps) at 300K, we find no bonds broken or geometric reconstructions in the final state. These results show NaV$_2$O$_4$ compound is also thermodynamically stable.

Furthermore, we can estimate the mechanical properties of NaV$_2$O$_4$ compound by using the stress-strain method. Tetragonal crystal structure totally possesses five independent elastic constants, which include C$_{11}$, C$_{12}$, C$_{13}$, C$_{33}$ and C$_{44}$. Other mechanical parameters such as the bulk modulus B, the Young's modulus E, the shear modulus G, and the Poisson's ratio $\nu$ can be derived from these elastic constants. As shown in Fig. 3, we can observe strong mechanical anisotropy for all the mechanical parameters. The mechanical stability of NaV$_2$O$_4$ compound can be estimated based on the generalized elastic stability criteria. For a stable tetragonal crystal structure, the following conditions are required:
\begin{equation} \label{eqn2}
  \begin{split}
C_{11} > |C_{12}|\\
2C_{13}^2 < C_{33}(C_{11} + C_{12})\\
C_{44} > 0\\
C_{11} - C_{12} > 0
  \end{split}
\end{equation}
By checking our results, we find all conditions in (1) are satisfied in NaV$_2$O$_4$ compound. This indicates that the material is mechanically stable.

\begin{figure}
\includegraphics[width=8.0cm]{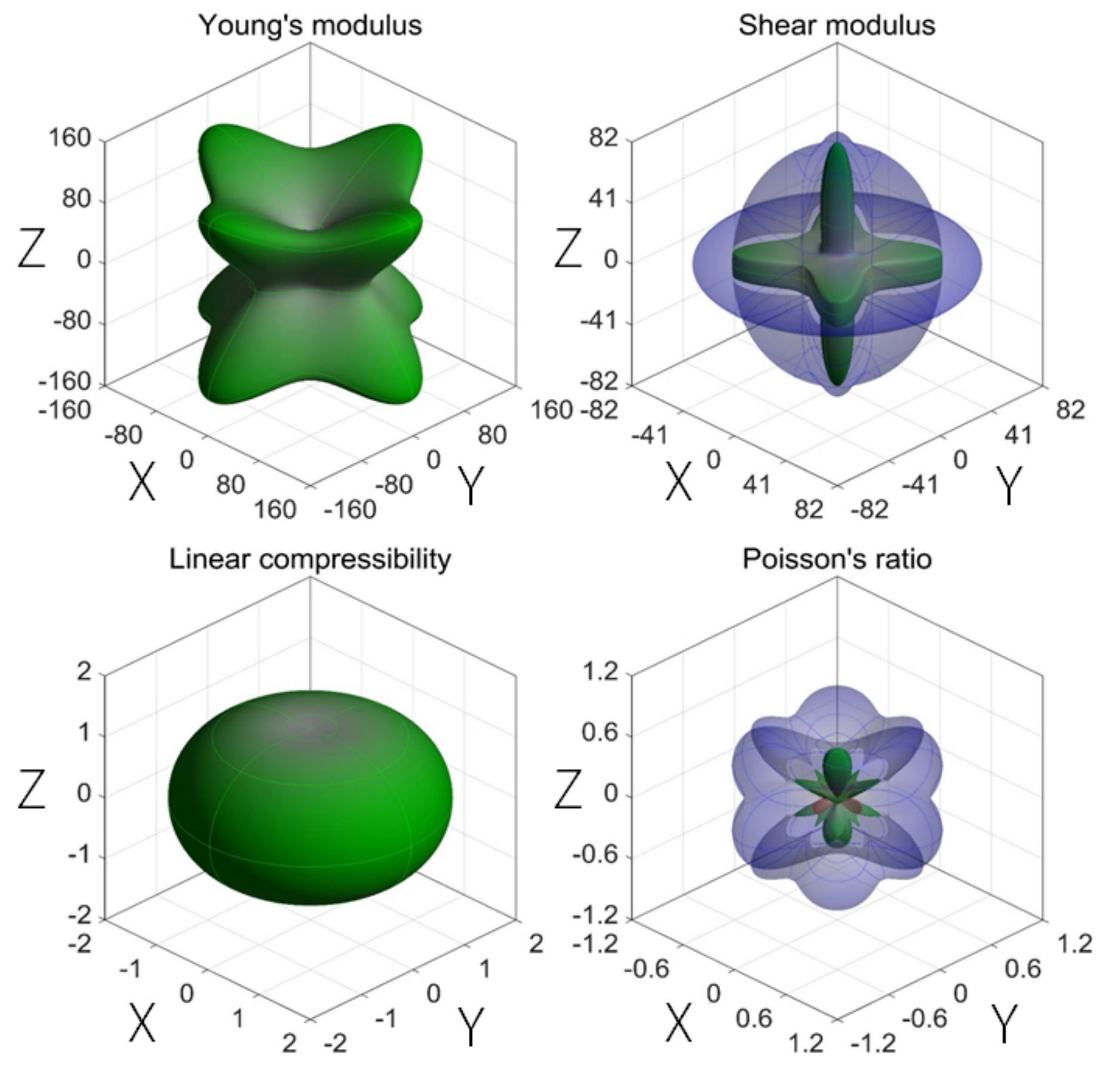}
\caption{The calculated directional dependent mechanical properties for NaV$_2$O$_4$. For the shear modulus and Poisson's ratio, the blue shaded surface indicates the maximum values and the green surface for the minimum ones.
\label{fig3}}
\end{figure}

Before study the electronic band structure, we need first determine the ground magnetic state in NaV$_2$O$_4$ compound. In NaV$_2$O$_4$, the magnetic moment is almost carried by the transition metal element V. To determine the ground magnetic state, we have compared the total energies with assigning the V moments in different magnetic configurations including nonmagnetic (NM), ferromagnetic (FM), and antiferromagnetic (AFM). Our results confirm that the FM state possesses the lowest energy, which is about 94.2 meV lower than the AFM state and 5.55 eV lower than the NM state for one unit cell. In the ground state, the magnetic moment is estimated to be 6 $\mu_B$ per unit cell. Moreover, the easy magnetization direction is determined to along the out-of-plane [001] direction, which is about 0.67 meV lower than typical in-plane directions.

\section{Weyl line and Weyl nodes without SOC}

\begin{figure}
\includegraphics[width=8.8cm]{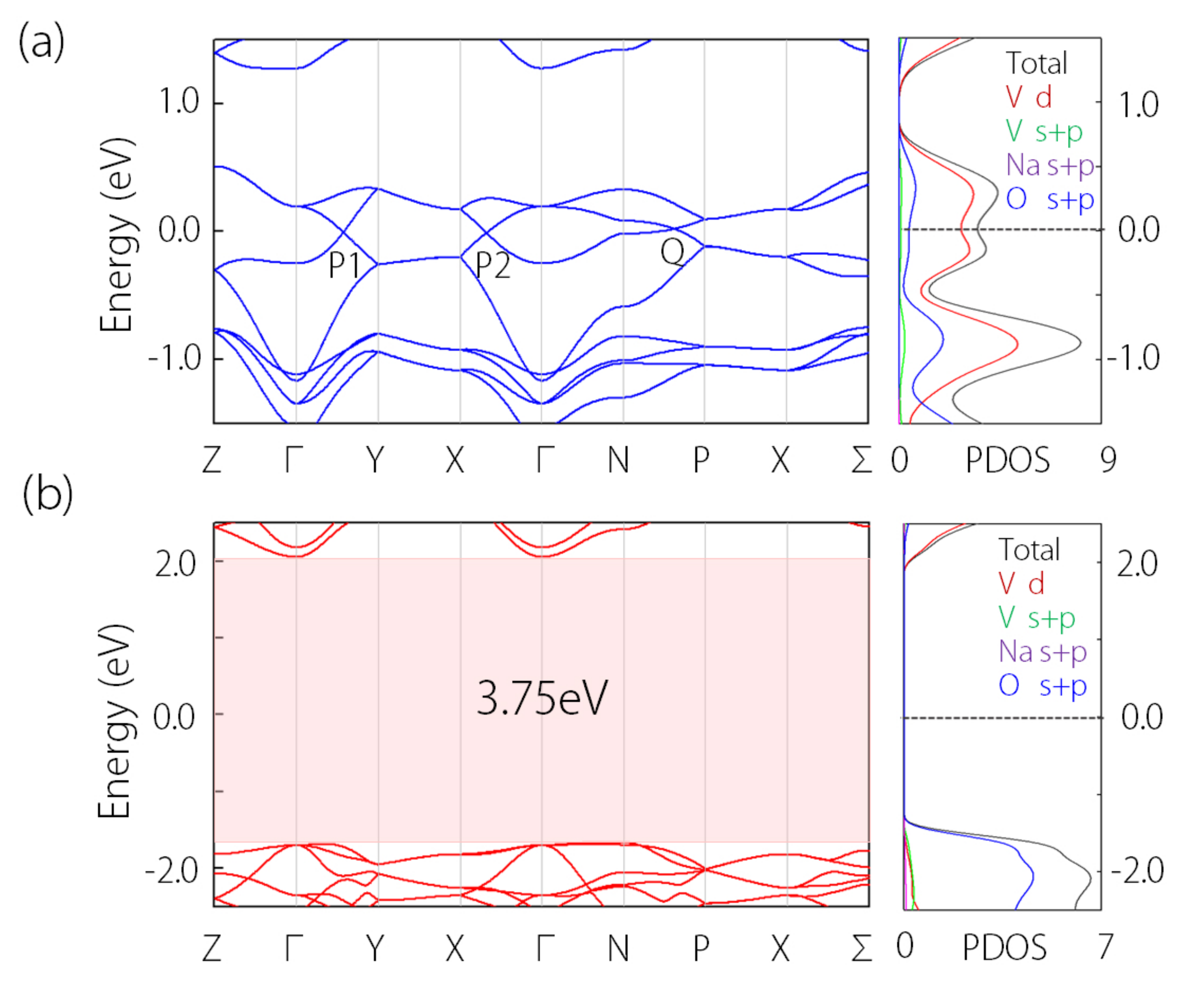}
\caption{Electronic band structure and the projected density of states of NaV$_2$O$_4$ in (a) spin up channel and (b) spin down channel. In (a), the band crossing points in the k-paths $\Gamma$-Y, X-$\Gamma$ and N-P are denoted as P1, P2 and Q, respectively. In (b), the size of band gap is labeled.
\label{fig4}}
\end{figure}

Based on the FM ground state, here we investigate the topological band structure of NaV$_2$O$_4$ compound. At first, the SOC is not included in the calculations. The spin-resolved band structure and the density of states (DOSs) are shown in Fig.~\ref{fig4}(a) and (b). We can observe that, the material shows a metallic band structure in the spin up channel. From the given total and partial DOSs, we find these states are mostly contributed by the V-d orbitals. Whereas, the material shows an insulating feature in the spin down channel, with the band gap as large as 3.75 eV. These results suggest that NaV$_2$O$_4$ is an excellent half metal with a 100$\%$ spin polarization of conducting electrons. Remarkably, we can observe three linear band crossing points (P1, P2, and Q) in the spin up bands [see Fig.~\ref{fig4}(a)], which occur in the $k$-paths $\Gamma$-Y, $\Gamma$-X, and N-P, respectively. All these band crossings locate quite close to the Fermi level.

\begin{figure}
\includegraphics[width=8.8cm]{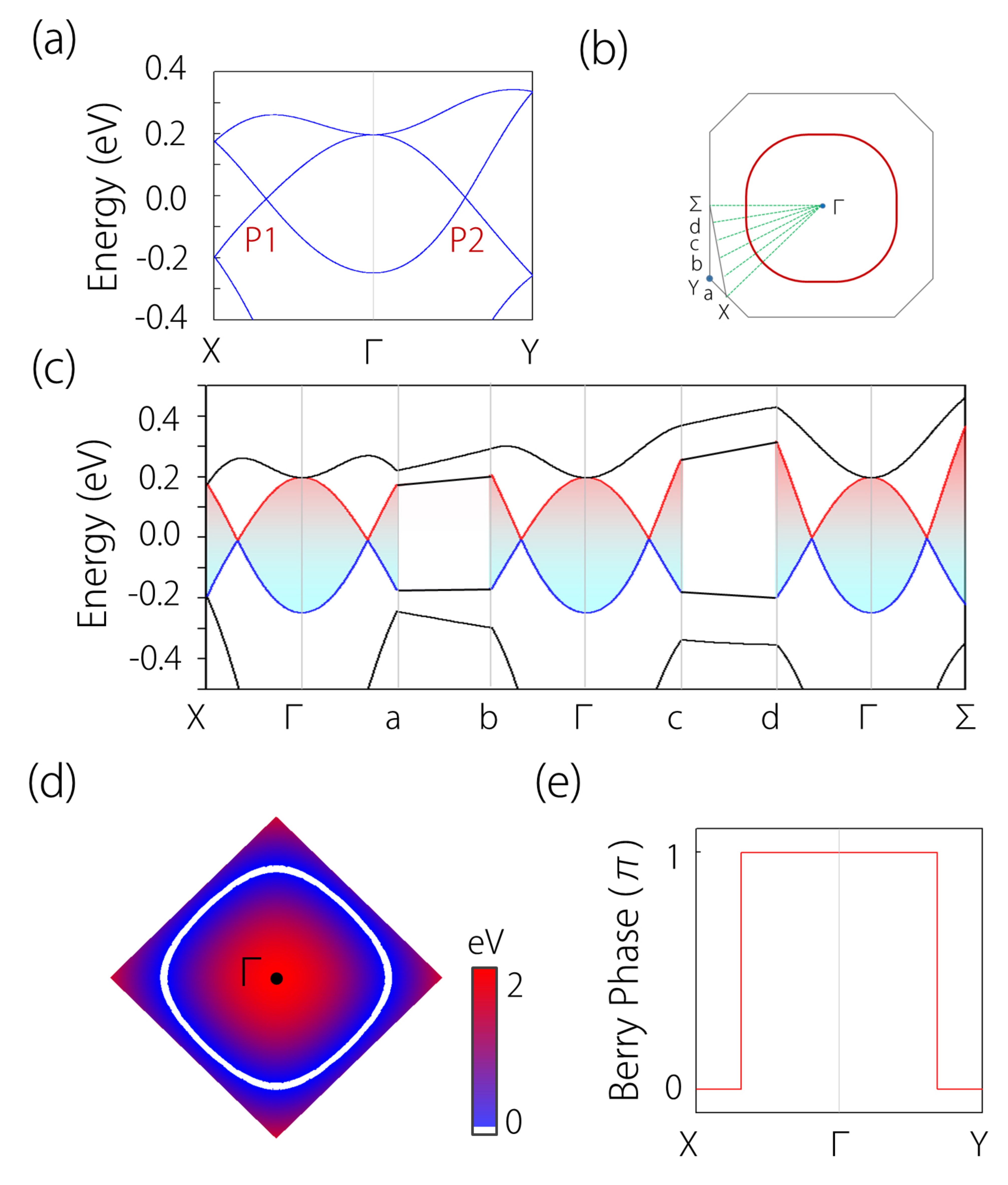}
\caption{(a) The enlarged view of band structure near P1 and P2 points. (b) Illustration of the Weyl nodal line in the $k_z$ = 0 plane and selected k-paths through the nodal line. (c) Band structure along the selected k-paths in (b). (d) The shape of nodal line in the Brillouin zone. (d) The Berry phase along the k-paths X-$\Gamma$-Y in the $k_z$ = 0 plane.
\label{fig5}}
\end{figure}

Firstly, we focus on the two band crossing points P1 and P2 in the spin up channel. The enlarged band structure near P1 and P2 is shown in Fig.~\ref{fig5}(a). It is worth noticing that, both the $k$-paths $\Gamma$-X and $\Gamma$-Y situate in the mirror-invariant plane $k_z$ = 0, which can protect a nodal line. To examine whether P1 and P2 are isolate nodal points or belong to a nodal line, we have calculated the band structure in several other $k$-paths, namely $\Gamma$-a, $\Gamma$-b, $\Gamma$-c, $\Gamma$-d and $\Gamma$-$\Sigma$ [see Fig.~\ref{fig5}(b)]. Together with $\Gamma$-X and $\Gamma$-Y, these $k$-paths can reflect the band signature of the whole Brillouin zone. The band structure on these $k$-paths is shown in Fig.~\ref{fig5}(c). It is clearly found that, all the $k$-paths $\Gamma$-a, $\Gamma$-b, $\Gamma$-c, $\Gamma$-d and $\Gamma$-$\Sigma$ show a linear band crossing point, suggesting the presence of nodal line in the $k_z$ = 0 plane. As shown in Fig. 5(c), the linear energy range near the nodal line is large with the size of about 0.5 eV. In Fig.~\ref{fig5}(d), we map the profile of the nodal line in the $k_z$ = 0 plane, it centers the $\Gamma$ point and has a square symmetry. The nodal line is protected by the mirror symmetry $M_z$, because the two crossing bands have opposite mirror eigenvalues (+1 and -1), as verified by our DFT calculations. To further show the nontrivial topology of the nodal line, we have calculated the Berry phase by integrating around a closed loop in the Brillouin zone. The result is shown in Fig.~\ref{fig5}(e). It clearly shows that the Berry phase is nontrivial.

\begin{figure}
\includegraphics[width=8.8cm]{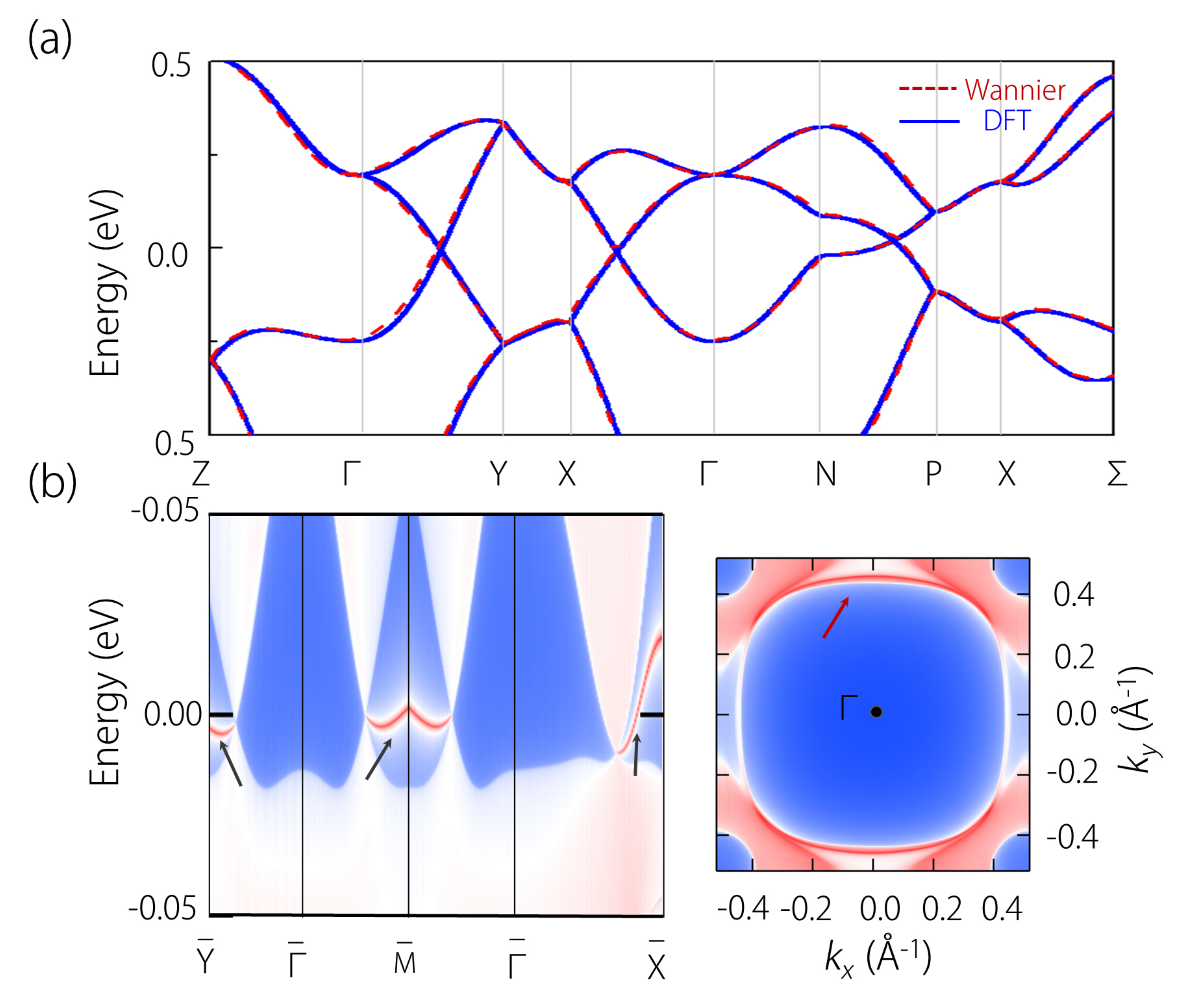}
\caption{(a) The comparison of band structure from the Wannier model and DFT. (b) Calculated (001) surface band structure with the drumhead surface states pointed by the black arrows. (c) The constant energy slice of the surface states at -3 meV.
\label{fig6}}
\end{figure}

Here, we develop an effective model for the nodal line. The crossing bands at the $\Gamma$ point respectively correspond to irreducible representations $A_{1g}$ and $A_{2g}$ for the $D_{4h}$ point group. Then the model respects the following symmetries: a fourfold rotation $C_{4z}$, a twofold rotation $C_{2z}$, a mirror $M_y$ and the inversion symmetry. Expanding up to the second order of \emph{k}, the effective Hamiltonian in low energy is given by
\begin{eqnarray}
  \mathcal{H} &=& [M+\alpha (k_x^2+k_y^2)+\beta k_z^2]\sigma_z+\gamma k_x k_y\sigma_x.
\end{eqnarray}
Here, $\sigma$'s is the Pauli matrix, and M, $\alpha,\ \beta,\ \gamma$ are real parameters which can be derived by fitting the DFT band structures. This effective Hamiltonian indicates a nodal loop lying on plane $k_z=0$.

Drumhead surface state is a distinguishing feature of nodal line materials~\cite{35,62,63,64,65,66,67}. To study the surface states of the nodal line in NaV$_2$O$_4$ compound, we first construct a tight-binding model by projecting onto the Wannier orbitals. The results are shown in Fig.~\ref{fig6}(a). Clearly, the Wanier band structure is in a good agreement with the band structure from the DFT. Correspondingly, the surface spectrum can be obtained. We show the (001) surface band structure in Fig.~\ref{fig6}(b). We can clearly observe the drumhead surface bands originating from the projected nodal line (as pointed by the black arrows). Besides, we also show a constant energy slice at 3 meV below the Fermi level. As shown in Fig.~\ref{fig6}(c), a closed surface loop centering the $\Gamma$ point is observed (as pointed by the red arrow). Such clear surface states greatly favor the experimental detection in future. To be noted, the drumhead surface states in NaV$_2$O$_4$ are also fully spin-polarized, being different with those in nonmagnetic nodal line materials.

\begin{figure}
\includegraphics[width=8.8cm]{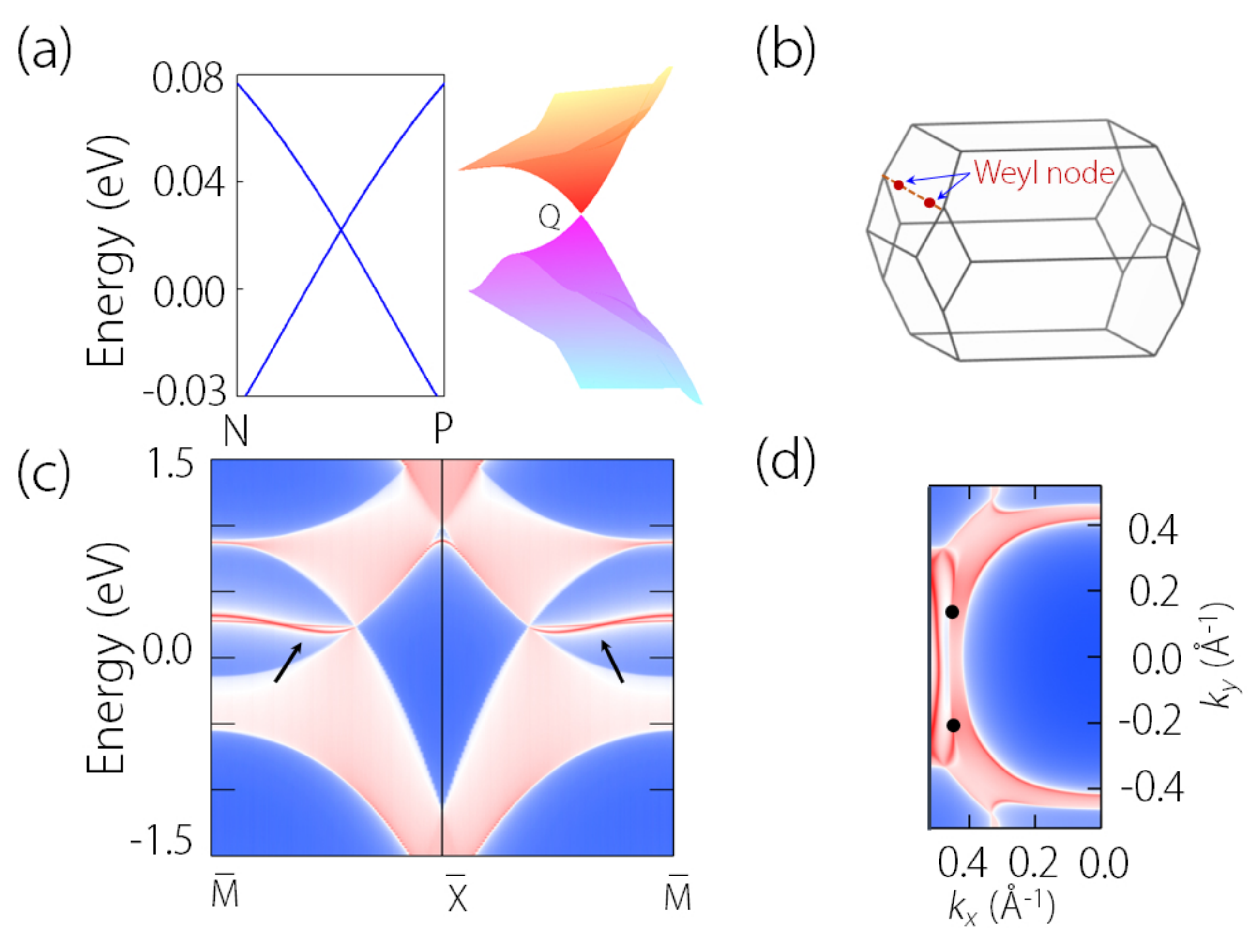}
\caption{(a) The enlarged view of band structure in the N-P path and the 3D plotting of band dispersion near Q point. (b) The position of the Q point in the BZ. (c) Calculated (001) surface band structure and (d) the corresponding constant energy slices at E= 0.02 eV.
\label{fig7}}
\end{figure}

Next, we focus on the linear band crossing labeled as Q in the N-P path. The enlarged view of band structure is shown in the left panel of Fig.~\ref{fig7}(a). The three dimensional plotting of band dispersion near Q is depicted in the right panel of Fig.~\ref{fig7}(a). We can clearly find that point Q is in fact a Weyl node. The Weyl node locates quite close to the Fermi level (0.02 eV). Considering the crystal symmetry, there totally exist four pairs of such Weyl nodes in the system. The positions for a pair of Weyl nodes in the Brillouin zone are shown in Fig.~\ref{fig7}(b). The projected spectrum on the (001) surface and the corresponding constant energy slice at 0.02 eV are shown in the Fig.~\ref{fig7}(c) and (d), respectively. We can clearly observe the Fermi arc surface states connecting the Weyl nodes.

\section{Effects of SOC, lattice strain and electron correlation}

\begin{figure}
\includegraphics[width=8.8cm]{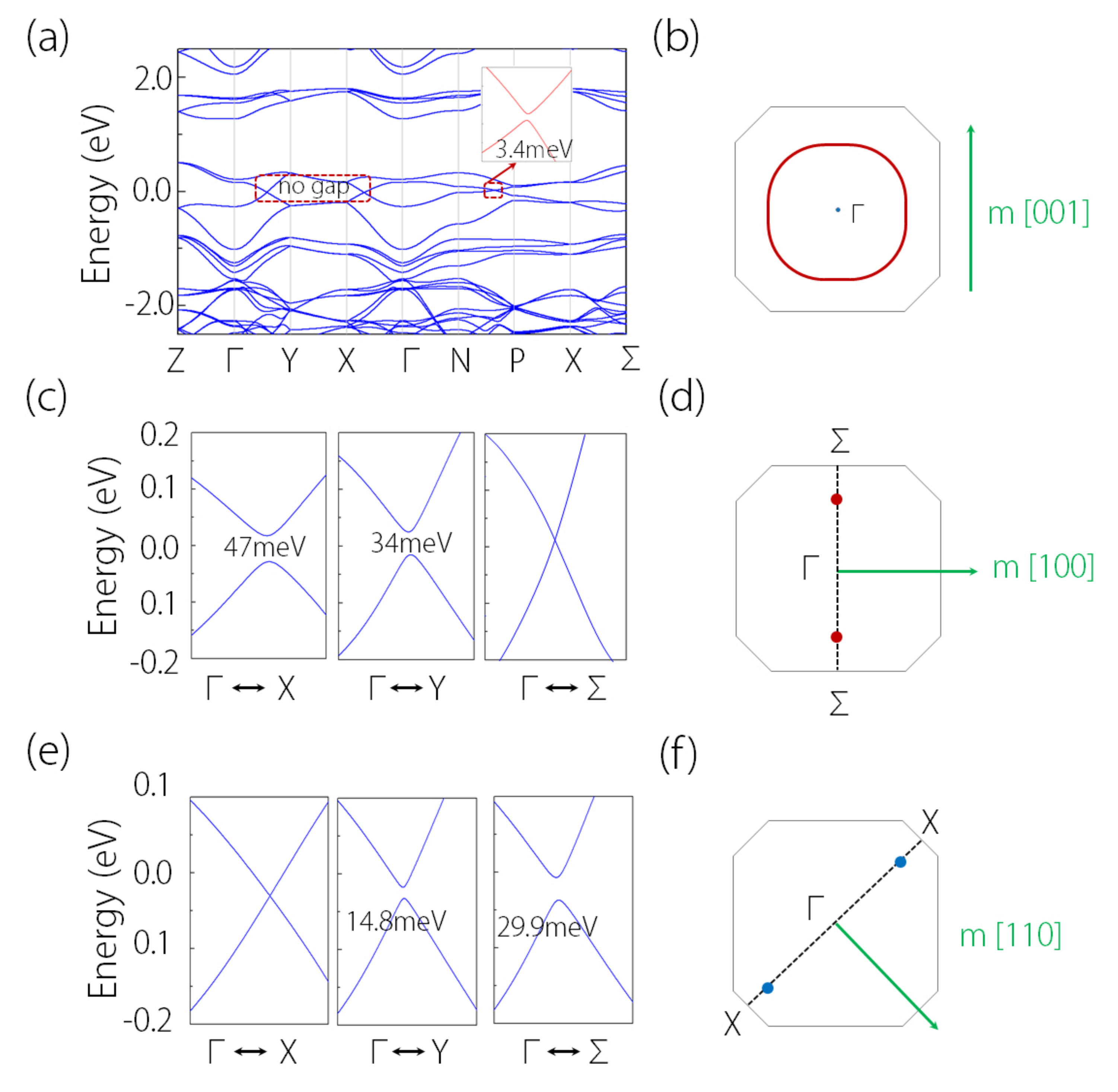}
\caption{(a) Electronic band structure of NaV$_2$O$_4$ under SOC with magnetization along the [001] direction. The inset of (a) shows the band gap in the N-P path. (b) Schematic diagram of the fully spin-polarization nodal line in $k_z$ = 0 plane under the [001] magnetization.
\label{fig8}}
\end{figure}

Above discussions are based on the electronic band structure without SOC. In NaV$_2$O$_4$ compound, the SOC effect cannot be simply ignored because the material has considerable atomic weight. Figure~\ref{fig8}(a)shows the band structure of NaV$_2$O$_4$ with SOC included. We find the bands in both spin channels conjunct together, but the band details do not change much. A careful examination on the band structure at crossing points P1, P2, and Q finds that the crossing points P1 and P2 are retained under SOC while Q is gapped. We have fully scanned the bands in the $k_z$ = 0 plane, and find the crossing along the whole nodal line is retained. These results suggest the nodal line in NaV$_2$O$_4$ compound is robust against SOC. This can be derived from the symmetry point of view. The material shows the easy magnetization along the out-of-plane [001] direction, and the mirror symmetry $M_z$ is preserved under SOC. Our calculations show that the two crossing bands for the nodal line still have opposite mirror eigenvalues (+i and -i), which protect the nodal line from opening gaps under SOC. As the results, NaV$_2$O$_4$ compound only shows a single Weyl nodal line in the $k_z$ = 0 plane when SOC is included, as shown in Fig.~\ref{fig8}(b).

The magnetic symmetry will change by shifting the magnetization direction, and the topological state may change accordingly. If in-plane magnetizations are applied in NaV$_2$O$_4$ compound, the mirror symmetry $M_z$ will be broken. In this occasion, the nodal line in the $k_z$ = 0 plane should vanish. In Fig.~\ref{fig8}(c) and (e), we show enlarged band structures with the magnetization along the in-plane [100] and [110] directions, respectively. For the [100] magnetization, we find the crossing points in the $\Gamma$-Y and $\Gamma$-X paths are gapped, but that in the $\Gamma$-$\Sigma$ path is retained [see Fig.~\ref{fig8}(c)]. We have also checked the bands for other parts on the nodal line, and no other crossing point is found. Therefore, NaV$_2$O$_4$ compound has transformed into a Weyl semimetal with a single pair of Weyl nodes in the $\Gamma$-$\Sigma$ path [see Fig.~\ref{fig8}(d)]. If the magnetization along the [110] direction, only the band crossing point in the $\Gamma$-X path is preserved [see Fig.~\ref{fig8}(e)]. As the results, the system exhibits a single pair of Weyl nodes in the $\Gamma$-X path [see Fig. 8(f)]. In fact, the system always retains a pair of Weyl nodes under arbitrary in-plane magnetization, which is ensured by symmetry. Here, we use the [110] magnetization as an example, while the symmetry analyses for other cases are fundamentally the same. Under the [110] magnetization, the mirror symmetry $M_z$ is broken, which gaps the nodal line. However, the system preserves the magnetic double point group $C_{2h}$. Along the $\Gamma$-$\Sigma$ path, the little group is $C_2$, which can allow presences of Weyl nodes in this path.

Above discussions have shown that NaV$_2$O$_4$ compound takes fully spin-polarized nodal line state in the easy magnetization direction and the state is robust against SOC. Here we further investigate the robustness of the fully spin-polarized line against lattice strain and the electron correlation effects. To ensure the fully spin-polarized nodal line state, two crucial conditions need to be satisfied: (i) an insulating band gap in the spin-up channel; (ii) band crossings the spin-up. These conditions ensure the half metal signature and the presences of nodal line.

\begin{figure}
\includegraphics[width=8.8 cm]{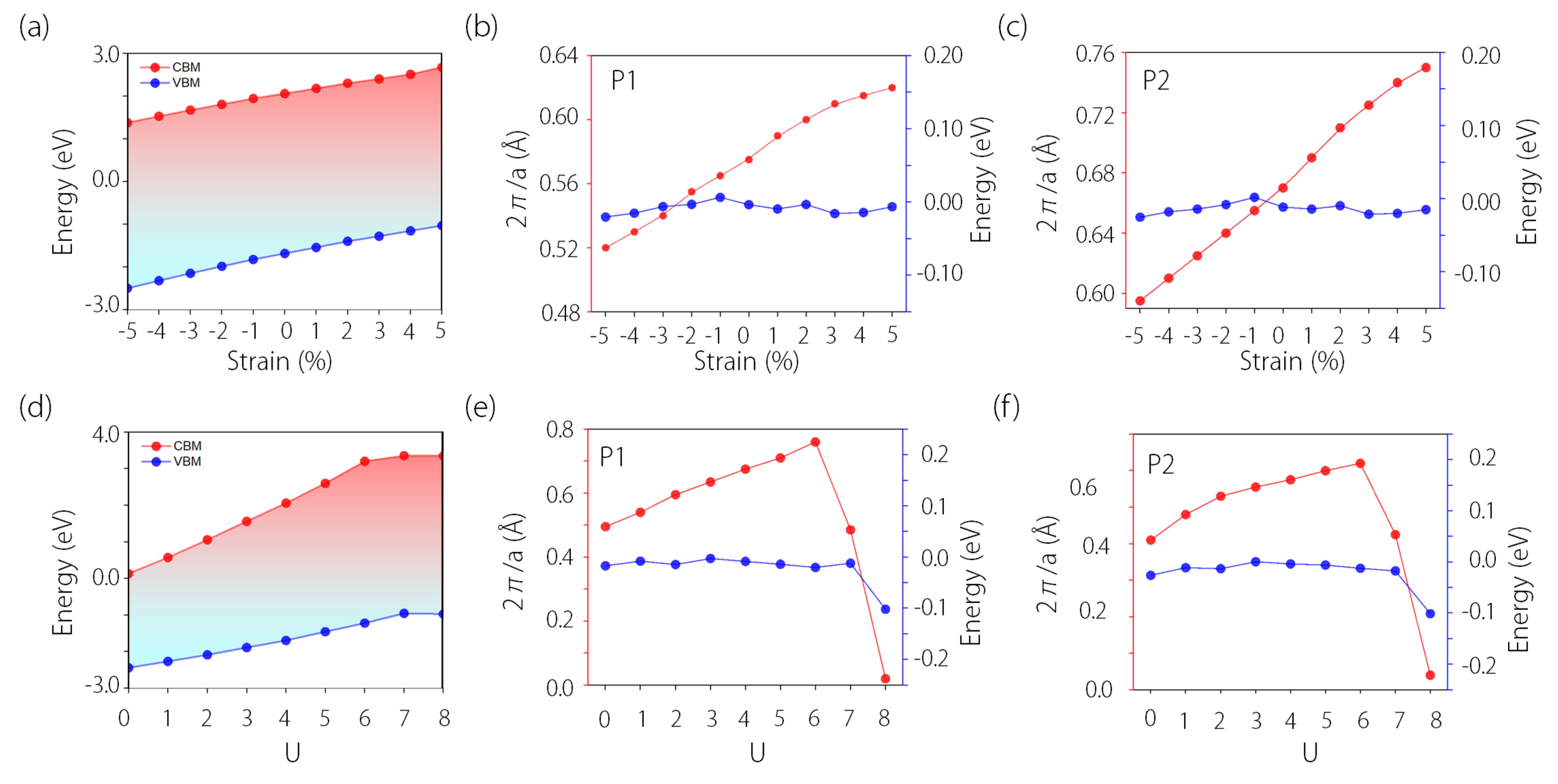}
\caption{(a) Illustration of the positions of valence band maximum (VBM) and conduction band minimum (CBM) in spin-down channel under strains. During the period, the change of the position and the energy for crossing points P1 and P2 are shown in (b) and (c). (d), (e), and (f) are similar with (a), (b), and (c), but for the case under different $U$ values of V atom.
\label{fig9}}
\end{figure}

 In Fig.~\ref{fig9}(a), we show the positions of the valence band maximum (VBM) and the conduction band minimum (CBM) under ¡À5$\%$ hydrostatic strains. Here "+" represents the tensile strain while "-" denotes the compressive strain. We can find that both the energies for VBM and CBM rise to higher energy levels with strain turning from compressive to tensile one. However, the insulating gap corresponding to the Fermi level always retains during the period. We have also checked the band structure in the spin-up channel, and found the crossing points P1 and P2 are always preserved. In Fig.~\ref{fig9}(b), we show the position of P1 in the momentum space under different strains. We can find that P1 moves away from the §¤ point with strain turning from compressive to tensile one. However, the energy of P1 does not change much. Similar phenomenon is also observed for P2, as shown in Fig.~\ref{fig9}(c). These results indicate that, the size of nodal loop in $\mathrm{NaV}_2$$\mathrm{O}_4$ compound would become larger under tensile strain but become smaller under compressive strain. In addition, we also investigate the robustness of the fully spin-polarized line state against the electron correlation effects. Here we shift the $U$ values of V atom from 0 to 8 eV. The results are shown in Fig.~\ref{fig9}(d)-(f). We find the fully spin-polarized line always preserves during the whole period. Above discussions have fully shown that the nodal line state in $\mathrm{NaV}_2$$\mathrm{O}_4$ compound is quite robust.

\section{Discussions and conclusion}

Before ending, we have two remarks. First, the nodal line in NaV$_2$O$_4$ compound is almost ideal for further experimental detections in that: (i) it locates slightly below the Fermi level, in the energy range of -8.7 meV to -1.5 meV; (ii) the nodal line band structure is fairly clean without other bands nearby; (iii) its linear energy range is as large as 0.5 eV; (iv) the nodal line is robust against SOC and lattice strain; (v) the drumhead surface states are very clear.

In addition, the nodal line state in NaV$_2$O$_4$ compound also shows following features: (i), it belongs to the type-I nodal line, because all the crossings around the line have the type-I band dispersion; (ii) considering the system lacks the time-reversal symmetry, the nodal line is a Weyl line with a double degeneracy; (iii) the nodal-line fermion and its corresponding drumhead surface states are fully spin-polarized, which are drastically different with those in nonmagnetic system.

In conclusion, we have revealed that NaV$_2$O$_4$ compound is a half metal with fully spin-polarized nodal line and tunable Weyl states. We find the material has excellent dynamical, thermal and mechanical stability. It shows a ferromagnetic ground state with the easy magnetization along the [001] direction. The spin-resolved band structure exhibits a metallic character in the spin up channel but an insulating one in the spin down channel. Two spin up bands cross with each other and form one nodal line and four pairs of Weyl nodes near the Fermi level. The nodal line and Weyl nodes have a 100$\%$ spin-polarization. Their nontrivial surface states are also spin-polarized, which have been clearly identified. We find the nodal line is robust against SOC, protected by mirror symmetry. We further find that the nodal line can shift into single pair of Weyl nodes by applying the magnetization in-plane. Moreover, the presence of fully spin-polarized nodal line in NaV$_2$O$_4$ is robust against lattice strain and the electron correlation effects. The fully spin-polarized nodal line state proposed here is very promising for spintronics applications.

\begin{acknowledgments}
This work is supported by National Natural Science Foundation of China (Grants No. 11904074), Nature Science Foundation of Hebei Province (No. E2019202222 and E2019202107). One of the authors (X.M. Zhang) acknowledges the financial support from Young Elite Scientists Sponsorship Program by Tianjin.
\end{acknowledgments}

\end{document}